\documentclass[a4paper]{spie}  
\usepackage[]{graphicx}
\usepackage[]{amssymb}
\usepackage{amsmath}
\usepackage[squaren,textstyle]{SIunits}
\addunit{\arcsec}{as}
\addunit{\parsec}{pc}

\title{The GRAVITY metrology system: narrow-angle astrometry via phase-shifting interferometry} 

\author{\large M. Lippa$^\ast$\supit{a}, N. Blind\supit{a}, S. Gillessen\supit{a}, Y. Kok\supit{a}, J. Weber\supit{a}, F. Eisenhauer\supit{a}, O. Pfuhl\supit{a}, \\A. Janssen\supit{a}, M. Haug\supit{a}, F. Hau\ss{}mann\supit{a}, S. Kellner\supit{a}, O. Hans\supit{a}, E. Wieprecht\supit{a}, T. Ott\supit{a}, \\L. Burtscher\supit{a}, R. Genzel\supit{a}, E. Sturm\supit{a}, R. Hofmann\supit{a}, S. Huber\supit{a}, D. Huber\supit{a}, S. Senftleben\supit{a}, \\A. Pfl\"uger\supit{a}, R. Gre\ss{}mann\supit{a}, G. Perrin\supit{b}, K. Perraut\supit{c}, W. Brandner\supit{d}, C. Straubmeier\supit{e}, \\A. Amorim\supit{f}, M. Sch\"oller\supit{g}  
\skiplinehalf
\small
\supit{a}Max Planck Institute for Extraterrestrial Physics (MPE), Giessenbachstr. 1, 85748 Garching, Germany; \\
\supit{b}Observatoire de Paris/LESIA 61, Av. de l'Observatoire, 75 014 Paris, France; \\
\supit{c}IPAG, 414 Rue de la Piscine, Domaine universitaire, 38 400 Saint Martin d'H\`eres, France; \\
\supit{d}MPIA Heidelberg, K\"onigstuhl 17, 69117 Heidelberg, Germany ; \\
\supit{e}Univ. Cologne, Z\"ulpicher Str. 77, 50937 K\"oln, Germany ; \\
\supit{f}SIM FCUL, Edif\'icio C8, gab 8.5.12, 1749-016 Lisboa, Portugal ; \\
\supit{g}ESO Garching, Karl-Schwarzschild-Str. 2, 85748 Garching, Germany}

\authorinfo{$^\ast$Further author information: \\ Magdalena Lippa, E-mail: lena@mpe.mpg.de}

\begin{document} 
  \maketitle 

\small Copyright 2014 Society of Photo-Optical Instrumentation Engineers. One print or electronic copy may be made for personal use only. Systematic reproduction and distribution, duplication of any material in this paper for a fee or for commercial purposes, or modification of the content of the paper are prohibited.
\normalsize

\begin{abstract}
The VLTI instrument GRAVITY will provide very powerful astrometry by combining the light from four telescopes for two objects simultaneously. It will measure the angular separation between the two astronomical objects to a precision of $\unit{10}{\micro\arcsec}$. This corresponds to a differential optical path difference (dOPD) between the targets of few nanometers and the paths within the interferometer have to be maintained stable to that level. For this purpose, the novel metrology system of GRAVITY will monitor the internal dOPDs by means of phase-shifting interferometry. We present the four-step phase-shifting concept of the metrology with emphasis on the method used for calibrating the phase shifts. The latter is based on a phase-step insensitive algorithm which unambiguously extracts phases in contrast to other methods that are strongly limited by non-linearities of the phase-shifting device. The main constraint of this algorithm is to introduce a robust ellipse fitting routine. Via this approach we are able to measure phase shifts in the laboratory with a typical accuracy of $\lambda/2000$ or $\unit{1}{\nano\meter}$ of the metrology wavelength.  
\end{abstract}


\keywords{GRAVITY, metrology, interferometry, VLTI, phase shifting, narrow-angle astrometry}

\section{INTRODUCTION}
\label{sec:intro}  

GRAVITY is a four-telescope beam combiner instrument for the Very Large Telescope Interferometer (VLTI) at the Paranal Observatory in Chile. The K-band light from four telescopes will be combined by GRAVITY for two objects simultaneously. By that the beam combiner will not only be capable of high resolution imaging with a beam width of $\unit{4}{\milli\arcsec}$ but will also provide narrow-angle astrometry to a precision of $\unit{10}{\micro\arcsec}$ - far beyond the current limits. 
\begin{figure}
   \begin{center}
   \begin{tabular}{c}
   \includegraphics[height=4cm]{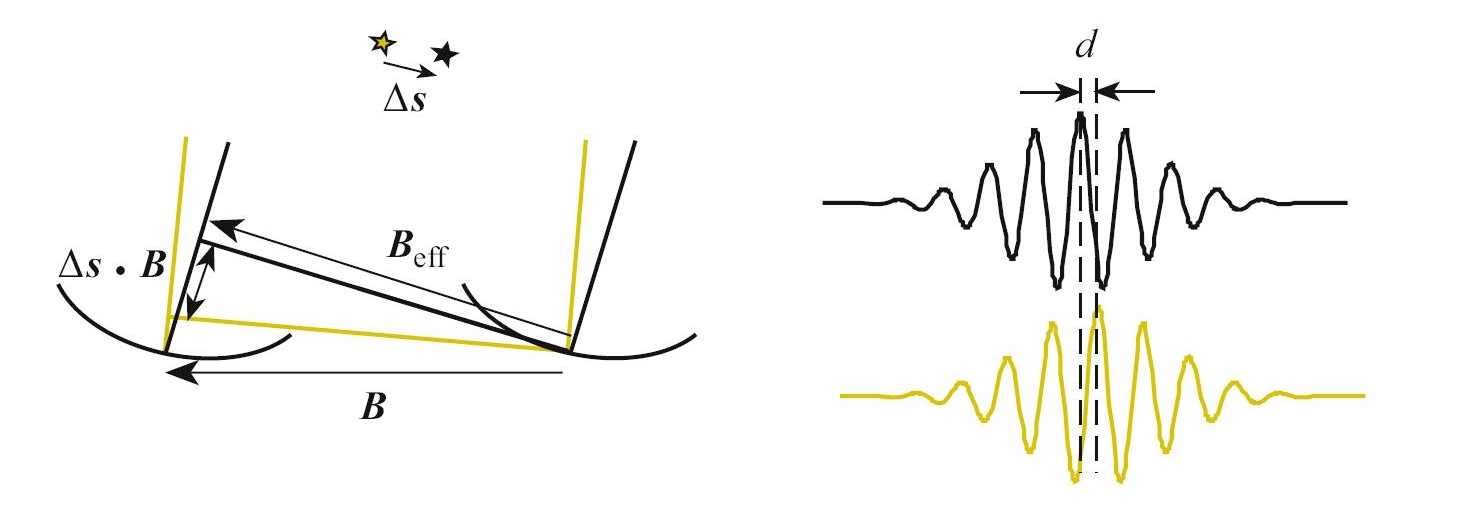}
   \end{tabular}
   \end{center}
   \caption[example] 
   { \label{fig:dOPD} 
Definition of dOPD. The differential optical path difference (dOPD) between two stars is the product $\Delta s \cdot B$ of the angular separation $\Delta s$ between the objects and the interferometer baseline $B$. This quantity can be derived by measuring the distance d between the white-light positions of the respective interference patterns of the targets when internal and atmospheric optical path differences are subtracted \cite{glindemann}.}
   \end{figure} 
One of the primary science goals targeted with these capabilities will be to probe physics in the Galactic Center close to the event horizon of the supermassive black hole. There, effects predicted by the theory of general relativity are expected to take place that cannot be observed in other, less extreme environments. Dynamical measurements taken with GRAVITY's astrometric mode are the key therefore. The idea is to observe two objects, one science and one reference target, simultaneously and to measure their angular separation by determining their so called Differential Optical Path Difference (dOPD) as illlustrated in Figure \ref{fig:dOPD}.

However, measuring the angular separation between the targets to a precision of $\unit{10}{\micro\arcsec}$ not only requires to measure their dOPD to a level of a few nanometers but also all paths within the interferometer from the telescopes down to the instrument need to be maintained stable to that level. Here is where the metrology system of GRAVITY comes into play which will monitor internal dOPDs by means of phase-shifting interferometry. In the following, the basic working principle of the GRAVITY metrology system is presented.

\section{The working principle of the GRAVITY metrology system} 
\subsection{Narrow-angle astrometry} 

\begin{figure}
   \begin{center}
   \begin{tabular}{c}
   \includegraphics[height=7cm]{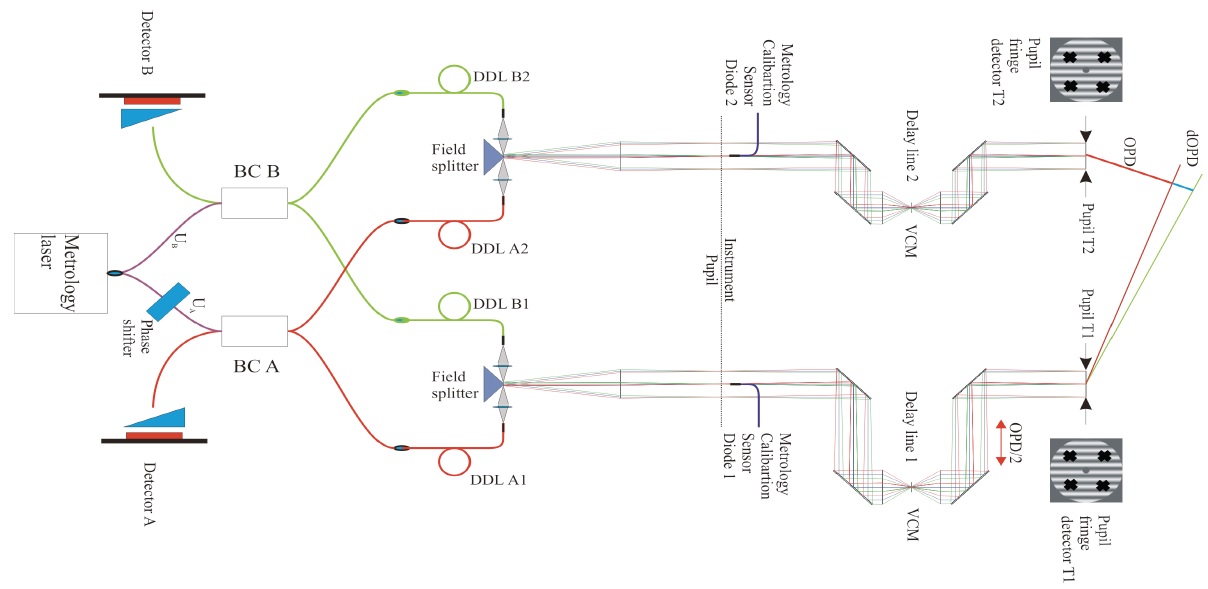}
   \end{tabular}
   \end{center}
   \caption[example] 
   { \label{fig:workingprinciple} 
The GRAVITY metrology system demonstrated in a schematic overview for two telescopes. A laser beam is launched which traces the optical paths from the GRAVITY beam combiners (BC A, BC B) backwards to the telescopes (T1, T2). There, fringe patterns are formed by the interfering metrology laser beams and sampled by four photodiodes per telescope. The fringes can be moved across the diodes by the phase-shifting device implemented in the chain at a kHz-rate. Via this method the phase information can be extracted from the measured intensities at the diodes that reconstruct the fringes \cite{gillessen12}.}
   \end{figure}

The underlying idea of the metrology system is to trace the optical paths of the science light within the interferometer by a laser beam, which is illustrated in Figure \ref{fig:workingprinciple}. The laser beam travels the paths in opposite directions being injected behind the beam combination facility in GRAVITY and going all the way backwards through GRAVITY and the VLTI infrastructures to the telescopes. The laser light at first is split into two equally bright beams, each of which is injected into one of the two beam combiners of the reference and science object. Before the injection a phase-shifting device is integrated in one of the beams such that in the end the interference patterns of the two beams at the telescope pupils can be sampled by phase-shifting interferometry.

For this reason a four-step phase-shifting algorithm is implemented in GRAVITY which applies four phase shifts A, B, C and D accompanied by the four corresponding intensity measurements at each individual telescope. This allows reconstructing the phases of the interfering laser beams. These phases correspond to the internal dOPDs which have to be subtracted from the respective phases of the science light beams measured at the GRAVITY detectors in order to obtain the true dOPD on sky between the targets. 

In more detail, a short mathematical derivation, considering only two telescopes for simplicity, can show that in such a design the dOPD needs to be calculated from three types of quantities, the science phases $\Psi_A$ and $\Psi_B$, measured at the detectors A and B, the metrology phases $\Phi_1$ and $\Phi_2$ at telescopes 1 and 2 as well as two terms related to the non-common path between metrology and science light, $\delta_1$ and $\delta_2$:
\begin{equation}
dOPD = \frac{\lambda_A}{2\pi}\Psi_A - \frac{\lambda_B}{2\pi}\Psi_B + \frac{\lambda_L}{2\pi} \left( \Phi_2-\Phi_1 \right) - \frac{\lambda_L}{2\pi} \left( \delta_2 - \delta_1 \right) \hspace{0.5cm} ,
\end{equation}
where $\lambda_A$ and $\lambda_B$ denote the effective wavelengths of the light of star A and B and $\lambda_L$ stands for the wavelength of the metrology laser.
Equivalent results are obtained when calculating the general case for four telescopes with six possible baselines. The difference of the non-common path terms $\delta_1$ and $\delta_2$ will be calibrated by swapping the light of two nearby stars between the beam combiners such that the zero points of the metrology phases are found. The actual adjustment will not be applied online but in the data reduction. Thus, the relevant quantities that have to be tracked by the metrology during observations are its phases at each telescope which correspond to internal dOPDs of the interferometer between the beam combination and the telescopes.

As already mentioned before, the phases of the metrology light are extracted by means of a four-step phase-shifting algorithm that shall be shortly introduced now. 

\subsection{Four-step phase-shifting algorithm} 
\label{sec:title}

The measurement of the internal dOPDs in the optical paths from the VLTI to the instrument GRAVITY is accomplished by determining the phases of fringe patterns that form in the pupil planes. The corresponding intensity patterns are sampled at a few points with photodiodes by applying phase shifts to one of the two metrology laser beams. Simultaneously, the resulting intensities of the shifted fringes are measured at the photodiodes, which encode the OPD information. A minimum of three phase shifts and intensity measurements respectively are necessary to determine the intensity pattern and by that the fringe phase. In this respect, the GRAVITY metrology uses four shifts $\alpha_i=\left\lbrace 0, \frac{\pi}{2}, \pi, \frac{3\pi}{2} \right\rbrace $ with $i\in\left\lbrace A,B,C,D\right\rbrace $ in order to determine the intensity distribution unambiguously denoted as ABCD algorithm.

According to the theory of interference of two beams with intensities $I_1$ and $I_2$,  the intensities measured for the individual phase steps should lie on the distribution $I(x,y,\alpha_i)$ as a function of position (x,y) on the detection area 
\begin{equation}
I(x,y,\alpha_i) = I_1 + I_2 + 2\sqrt{I_1 I_2}\sin(\Phi(x,y) + \alpha_i) \hspace{0.5cm} ,
\label{eq:fringes}
\end{equation} 

where $\Phi(x,y)$ denotes the fringe phase.

These four shifts and intensities can be written as four such equations with three unknown variables $I_1$, $I_2$ and $\Phi$. Solving this system of equations for the variable of interest, the fringe phase $\Phi$, leads to the simple relation below if the phase shifts correspond to exact multiples of $\frac{\pi}{2}$: 
\begin{equation}
\Phi=\arctan\left( \frac{I_A - I_C}{I_B- I_D}\right)  \hspace{0.5cm} .
\label{eq:phi}
\end{equation} 

In case the phase shifts $\alpha_i$ deviate from the nominal values this formula can be generalized if the deviations are known, but it is foreseen to calibrate the nominal shifts to extract the metrology phases \cite{sahlmann}.

The uncertainties in the phase extraction of the metrology enters the total astrometric error budget of GRAVITY, which is shown in the next section.

\subsection{Astrometric error budget} 

As raised above, the simple four-step algorithm implemented in the metrology will be based on exact multiples of $\frac{\pi}{2}$ being applied as phase shifts. This requires a calibration of the phase-shifting device. As a consequence, the precision of this calibration enters the uncertainty of the phase measurement. Following from the error propagation of Equation \ref{eq:phi} the phase uncertainty squared $\delta \Phi ^2$ amounts to:
\begin{eqnarray}
\delta \Phi ^2 &=& \left( \frac{\partial \Phi}{\partial I_A} \delta I_A\right)^2  + \left( \frac{\partial \Phi}{\partial I_B} \delta I_B\right)^2 + \left( \frac{\partial \Phi}{\partial I_C} \delta I_C\right)^2 +\left( \frac{\partial \Phi}{\partial I_D} \delta I_D\right)^2 \\
         &=& \frac{\delta I^2}{2I_0^2} \hspace{0.5cm} ,
\label{eq:difference}
\end{eqnarray}
assuming that the four intensity measurements have equal uncertainties $\delta I:=\delta I_A=\delta I_B=\delta I_C=\delta I_D$ and $I_1=I_2=I_0$.

Thus, the phase uncertainty in radian is defined by the reciprocal of $\sqrt{2}$ times the signal-to-noise ratio $I_0/\delta I$. For the shortest baselines the astrometric error budget leaves roughly a few nanometers to the phase measurement accuracy within 5 minutes, such that all individual uncertainties introduced by the devices involved in this measurement should amount to fractions of this value including the error coming from the calibration of the phase shifter. Table \ref{tab:astrometric} demonstrates the influence of the phase shifter calibration errors of order $\lambda/4000$, $\lambda/2000$ and $\lambda/1000$ on the astrometric error budget. The goal therefore is to calibrate the phase shifter to a nm-level of accuracy or even lower in order to reach the specified astrometric precision.

Before the details of developing an appropriate calibration routine are presented, the properties of the phase-shifting device used for the GRAVITY metrology are summarized.
\begin{table}[h]
\centering
\begin{tabular}{l|l|l|l}
\textbf{Error term} & $\boldsymbol{\lambda/4000}$ & $\boldsymbol{\lambda/2000}$ & $\boldsymbol{\lambda/1000}$ \\ \hline
Phase shifter calibration error & \unit{0.5}{\nano\meter} & \unit{1}{\nano\meter} & \unit{2}{\nano\meter} \\ \hline
Total metrology phase extraction error & \unit{0.94}{\nano\meter} & \unit{1.22}{\nano\meter} & \unit{1.81}{\nano\meter} \\ \hline
Total GRAVITY OPD error & \unit{6.83}{\nano\meter} & \unit{6.87}{\nano\meter} & \unit{7.00}{\nano\meter} \\ \hline
Astrometric error  & \unit{10.27}{\micro\arcsec} & \unit{10.34}{\micro\arcsec} & \unit{10.53}{\micro\arcsec} \vspace{0.3cm}
\end{tabular}
 \caption[Influence of the phase shifter calibration error on the astrometric error of GRAVITY]{Influence of the phase shifter calibration error on the astrometric error of GRAVITY. The numbers correspond to the observation case with four VLTI Unit Telescopes (UTs) during a period of \unit{5}{\minute}. The phase shifter calibration error is one of several terms, such as the performance of the metrology laser and receivers, that enters the total metrology phase extraction error. The latter also adds up quadratically with other errors to a total GRAVITY OPD error term which can be translated into an astrometric error. A phase shifter calibration error of \unit{2}{\nano\meter} would dominate the total phase extraction error of the metrology which then would become the second largest contribution to the astrometric error.}
\label{tab:astrometric}
\end{table}

\section{Calibration of the phase shifter} 

\subsection{Phase shifter} 

\begin{figure}
   \begin{center}
   \begin{tabular}{c}
   \includegraphics[height=4cm]{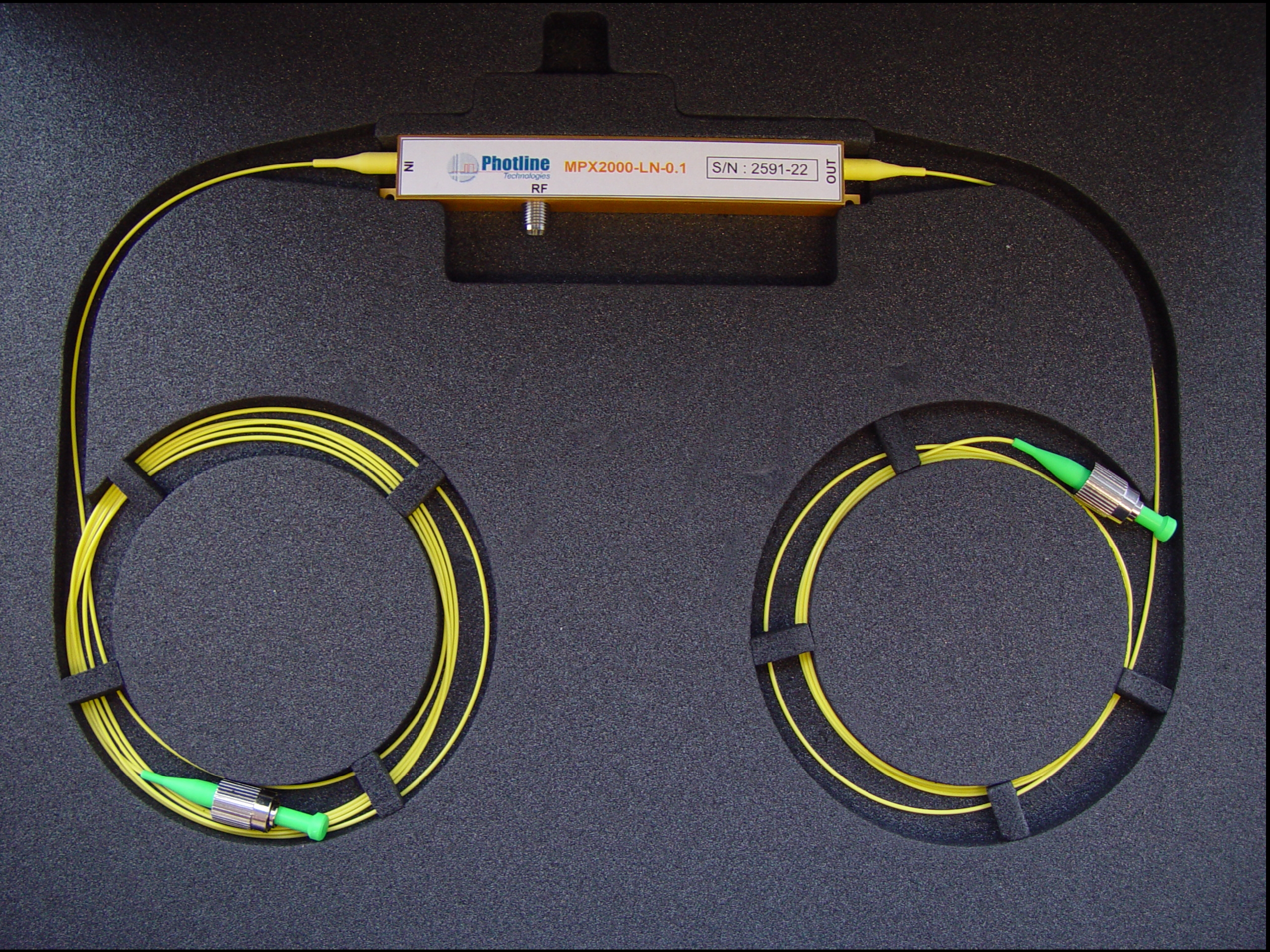}
   \end{tabular}
   \end{center}
   \caption[example] 
   { \label{fig:phaseshifter} Photline phase shifter of type MPX2000-LN-0.1 with input fiber to the left and output fiber to the right. The phase modulation is realized by setting a voltage to the radiofrequency input at the bottom of the phase shifter housing.
}
   \end{figure}

Our phase-shifting device of type MPX2000-LN-0.1 displayed in Figure \ref{fig:phaseshifter} is a component from Photline Technologies. The phase modulation of optical signals is based on a birefringent lithium niobate crystal (LiNbO$_3$). The refractive index of this material can be varied in proportion to the strength of an applied electric field. Thus, the phase of an electro-magnetic wave passing the crystal can be influenced by setting a voltage. On the surface of the crystal a waveguide is implemented by titanium diffusion. This diffusion zone leads to an increase of the refractive index such that total internal reflection keeps the injected light within the waveguide \cite{wooten}. The manufacturer's specifications of the device are given in Table \ref{tab:specs} together with the corresponding requirements from the GRAVITY metrology design.

\begin{table}[h]
\vspace{0.5cm}
\centering
\begin{tabular}{l|l|l}
\textbf{Parameter} & \textbf{Photline specification} & \textbf{Metrology requirement} \\ \hline
Operating wavelength & \unit{1900}{\nano\meter} -- \unit{2200}{\nano\meter} & \unit{1908}{\nano\meter}  \\ \hline 
Electro-optic bandwidth & \unit{150}{\mega\hertz}  & \unit{4}{\kilo\hertz}  \\ \hline
RF input power & \unit{-20}{\volt} -- \unit{+20}{\volt}  & \unit{0}{\volt} -- \unit{10}{\volt}  \\ \hline
Optical input power & \unit{0.1}{\watt}  & \unit{0.85}{\watt} 
\vspace{0.3cm}
\end{tabular}
 \caption[Specifications of the phase shifter]{Specifications of the phase shifter. The device is designed for wavelengths between $\unit{1900}{\nano\meter}$ and $\unit{2200}{\nano\meter}$. The electro-optic bandwidth of the phase shifter amounts to $\unit{150}{\mega\hertz}$ and the modulation range is specified from $\unit{-20}{\volt}$ to $\unit{20}{\volt}$. For the GRAVITY metrology roughly $\unit{0}{\volt}$ to $\unit{10}{\volt}$ are used which correspond to a full wave for a wavelength of $\unit{2000}{\nano\meter}$ and a bandwidth of $\unit{4}{\kilo\hertz}$ is foreseen to operate system. Long-term tests showed that the injected power level of $\unit{0.85}{\watt}$ is withstood by the component although it is specified for a maximum input of $\unit{0.1}{\watt}$ by the manufacturer Photline Technologies.}
\label{tab:specs}
\end{table}

The birefringent crystal has two transmission axes with different refractive indices. Since the metrology laser light is polarized and guided in PM fibers, the polarization can be maintained in this respect when being aligned to one of these axes. Due to the presence of several connectors in the metrology fiber chain small misalignments can occur. For this reason, we spliced fiber polarizers to the input and output of the phase shifter to correct for these mismatches. 
\newpage
The phase-shifting concept of the metrology system used to determine internal dOPDs in the GRAVITY interferometer requires that the accuracy of the applied phase shifts should be of order $\unit{1}{\nano\meter}$ or less. In practical terms, this means that the translation between the voltage applied to the phase shifter and the resulting phase shift has to be known to that level. This relation can be measured in an interferometric test setup. In order to determine the relation between the applied voltage and resulting phase shift we tested two methods, a linear scan of a full wave and a phase-step insensitive algorithm, both of which will be presented here.

\subsection{Linear scan} 

\begin{figure}
   \begin{center}
   \begin{tabular}{c}
   \includegraphics[height=7cm]{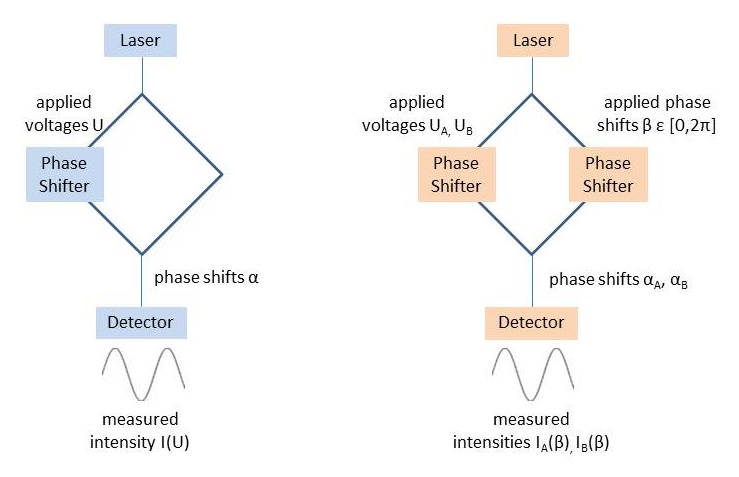}
   \end{tabular}
   \end{center}
   \caption[example] 
   { \label{fig:schemes} Two calibration schemes for the phase shifter. On the left side a simple interferometric setup with the phase shifter in one of the two interfering beams is displayed. With linear voltage ramps fringes are scanned by at least a full wave and their intensity distribution is measured by a detector as a function of the applied voltage U in order to determine the phase shifts $\alpha$ that correspond to the ABCD steps. On the right, another setup is shown which includes two phase shifters. One phase shifter is used as a delay line which is continuously shifted by many different phase shifts $\beta$ at a slow rate. For every phase shift $\beta$ a phase step $|\alpha_A-\alpha_B|$ is applied to the other phase shifter by setting $U_A$ and $U_B$. This phase step can be calibrated by measuring the corresponding intensities $I_A$ and $I_B$ as a function of $\beta$.
}
   \end{figure}

In a first simple approach, the main idea was to scan the range of a full wave by setting a linear voltage ramp from $\unit{0}{\volt}$ to $\unit{10}{\volt}$ in an interferometric test setup, which is drawn on the left side of Figure \ref{fig:schemes}. As in the final metrology design, the laser light is split into two channels of which one is feeding the phase shifter. Then the beams are combined again and the interferometric signal is read out with a receiver using the same type of photodiodes as our final design. 

\begin{figure}
   \begin{center}
   \begin{tabular}{c}
   \includegraphics[height=5cm]{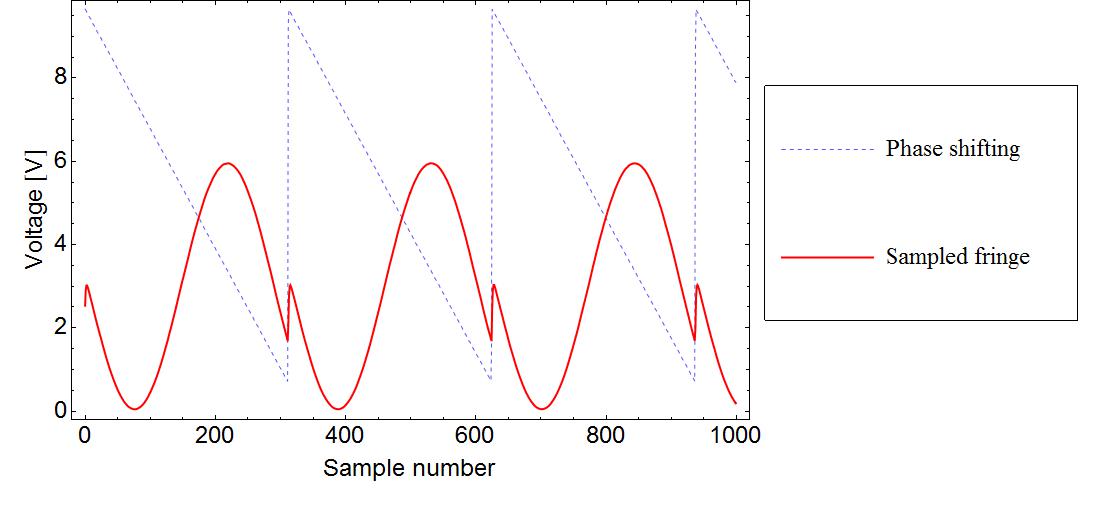}
   \end{tabular}
   \end{center}
   \caption[example] 
   { \label{fig:measurement} Linear scans of a full-wave modulation. The sawtooth voltage (dashed blue line) is applied to the phase shifter while the sinusoidal function (red) corresponds to the measured fringes at the receiver.
}
   \end{figure}

We average over a few hundred linear scans as in Figure \ref{fig:measurement}, at the maximum rate of $\unit{500}{\hertz}$ given by the response of our receiver design, to extract the intensity distribution of the interference pattern. Fitting a function of the following shape to the acquired data allows to model the desired relation between voltage U and phase shift $\alpha$:
\begin{equation}
I(U) = \xi + I_1 + I_2(U) + 2\nu\sqrt{I_1 I_2(U)}\sin(\Phi + \alpha(U)) \hspace{0.5cm} .
\label{eq:fringe}
\end{equation} 

The quantities $I_1$ and $I_2(U)$ correspond to the transmission functions of the two interfering beams. The intensity $I_1$ denotes the transmission of the unmodulated arm. The intensity $I_2$ is the transmission of the phase-shifted arm. It turns out that the transmission of the phase shifter depends on the applied voltage, i. e. is not independent of the applied phase step. We measure $I_1$ and $I_2(U)$ with linear scans by disconnecting one of the interfering fiber chains before we do the actual fringe measurement. $\xi$ and $\nu$ are parameters which we have to introduce to adjust the bias and contrast of $I$ which are not perfectly reconstructed by the previously measured $I_1$ and $I_2(U)$.
\newpage
If the relation $\alpha(U)$ of the phase shifter is linear, the intensity is a simple sine function and the four required voltages for the ABCD phase shifts are therefore easy to derive. However, phase-shifting devices can show non-linearities like the Kerr effect such that the function $\alpha(U)$ is of more complicated shape as in our case where a polynomial of 4th order is needed to fit the measured intensity distribution. This and other effects observed during the measurements strongly limit the achievable accuracy of the phase shifter calibration error and make the related analysis of the measured data complicated. 

\begin{figure}
   \begin{center}
   \begin{tabular}{c}
   \includegraphics[height=5.5cm]{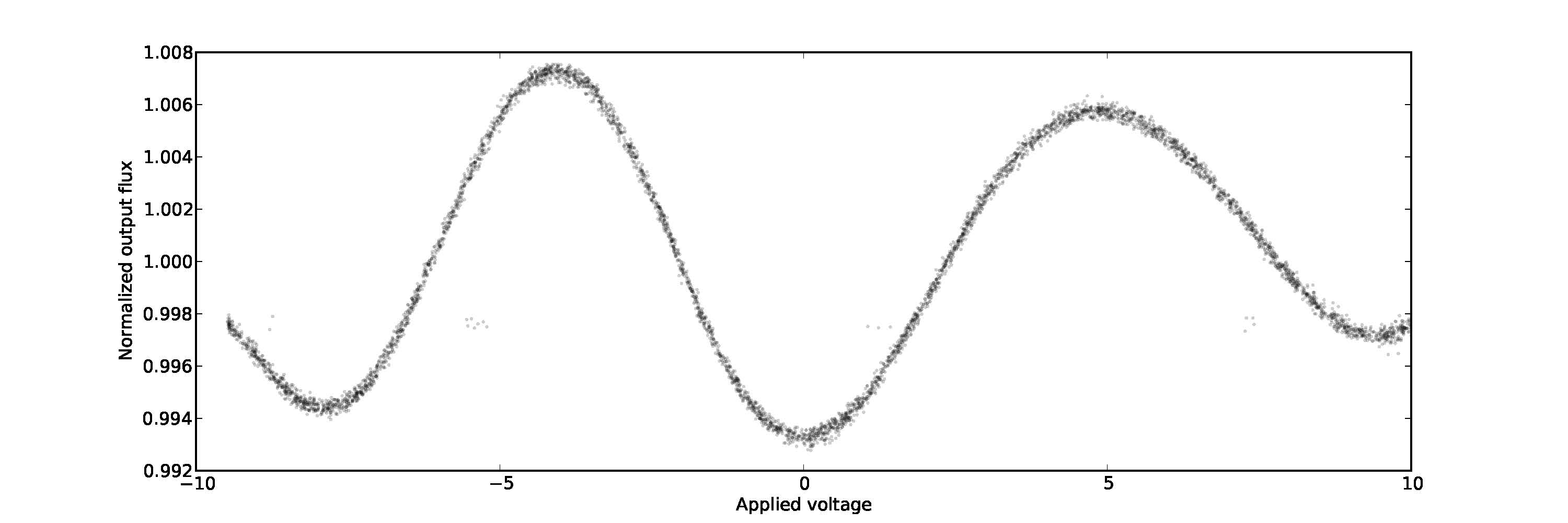}
   \end{tabular}
   \end{center}
   \caption[example] 
   { \label{fig:transmission} Measurement of the transmission modulation in the phase shifter. The corresponding setup consists of the laser source feeding the phase-shifting device and a receiver measuring its output intensity while voltage ramps from \unit{-10}{\volt} to \unit{+10}{\volt} were applied. The plot shows the averaged and normalized intensity distribution versus the applied voltage.
}
   \end{figure}

As a summary the limiting factors in our measurements were the following:
\begin{itemize}
\item The relation $\alpha(U)$ between the voltage U applied to the phase shifter and the resulting phase shift $\alpha$ is non-linear.
\item The measurements are influenced by laser power fluctuations of a few percent at frequencies from $\unit{0}{\hertz}$ to $\unit{50}{\hertz}$. 
\item The transmission of the phase shifter depends on the applied voltage with a peak-to-peak modulation of around $10^{-2}$ as shown in Figure \ref{fig:transmission} probably due to low finesse Fabry-Perot interference of Fresnel reflections in the phase shifter.
\item Fringe drifts of $\unit{0.2}{\nano\meter}$ per full wave occur due to environmental instabilities like temperature fluctuations and vibrations, which add up in the complete measurement of several hundred periods.
\item The birefringence of the phase shifter generates intrinsic contrast modulation due to polarization misalignments in the component.
\end{itemize}

These effects need to be taken into account when modeling the fitting function used for the measured intensity distribution. In this approach, we were able to determine phase shifts with a typical accuracy of $\unit{2}{\nano\meter}$ correcting for the non-linear transmission function and the fringe drifts, however under the assumption that our model function $I(U)$ and in particular $\alpha(U)$ describe reality well. Thus, our main concerns were:
\begin{itemize}
\item Does our model describe reality well enough?
\item Are the four ABCD values that we determine from this calibration scheme still correlated regarding that they are not measured instantaneously and in particular suffer from environmental instabilities and laser power fluctuations?  
\end{itemize}

These results caused us to investigate other calibration methods which are less affected by non-linearities. We were able to overcome the main drawbacks mentioned above by using a phase-step insensitive algorithm discussed in the next section.

\subsection{Phase-step insensitive algorithm} 

The basic idea behind the phase-step insensitive approach is to shift by the four values instantaneously that we believe to be close to the required ABCD phase shifts, namely $\left\lbrace\unit{0}{\volt},\unit{2}{\volt},\unit{4}{\volt},\unit{6}{\volt}\right\rbrace$ since a full wave is scanned by approximately $\unit{8}{\volt}$ for our wavelength. In this manner, we directly probe the phase shifts and the phase shifter non-linearities as well as environmental instabilities can be neglected, such that a fixed correlation is kept between the different steps or voltages applied to the component. With the phase shifter in one arm of the setup and a delay line in the other, as shown on the right in Figure \ref{fig:schemes}, we used this fact to determine our desired ABCD steps with a phase-step insensitive method. The delay line can be used to slowly move the phase in order to scan the fringes while the phase shifter is constantly repeating approximate ABCD steps for each delay line shift. 

The correlation of the instantaneous ABCD shifts remains fixed independently of the delay line phase. In this manner, one obtains four intensity distributions $I_i$ with $i\in\left\lbrace A,B,C,D\right\rbrace $ as a function of the delay line phase $\beta$:
\begin{equation}
I_i(\beta) = a_i + b_i\sin(\beta + \alpha_i) \label{eq:1} \hspace{0.5cm} ,
\end{equation} 
where the parameters $a_i$ correspond to the intensity offsets and $b_i$ to the amplitudes of the intensity modulation.

These functions have the form of Cartesian coordinates of an ellipse in its parametric form. Plotting couples of these intensity distributions, $I_i$ and $I_j$, against each other will result in an elliptic figure with a shape depending on different parameters of the system such as flux and phase shift $\Delta_{ij}=\alpha_i-\alpha_j$. The data can be fitted by the ellipse equation in Cartesian coordinates
\begin{equation}
EI_i^2+FI_iI_j+GI_j^2+HI_i+KI_j+L=0
\label{eq:ellipse}
\end{equation} 

with the coefficients (E, F, G, H, K, L), which allows for calculating the phase step $\Delta_{ij}$ by the following formula \cite{farrell}:
\begin{equation}
\Delta_{ij}=\arccos\left(\frac{-F}{\sqrt{4EG}}\right) \hspace{0.5cm} .
\label{eq:delta}
\end{equation} 

The parameters (E, F, G, H, K, L) can be written as
\begin{align}
\kappa &=(a_ib_j)^2+(a_jb_i)^2-2a_ib_ia_jb_j\cos(\Delta_{ij}) -b_i^2b_j^2\sin^2(\Delta_{ij})  \label{eq:kappa} \\
E &=b_j^2/\kappa \label{eq:a} \\
F &=-2b_ib_j\cos(\Delta_{ij})/\kappa  \label{eq:b} \\
G &=b_i^2/\kappa \label{eq:c} \\
H &=2(a_jb_ib_j\cos(\Delta_{ij})-a_ib_j^2)/\kappa \label{eq:d} \\
K &=2(a_ib_ib_j\cos(\Delta_{ij})-a_jb_i^2)/\kappa \label{eq:e} \\
L &=1 \hspace{0.5cm} .\label{eq:f}
\end{align} 

L is fixed to be 1 in order to avoid that the parameters shrink to zero in the $\chi^2$-minimization that we run in order to find the parameters $a_i$, $a_j$, $b_i$, $b_j$ and $\Delta_{ij}$ such that equation \ref{eq:ellipse} tends to zero. It is also possible to fit the coefficients (E, F, G, H, K, L) by direct analytical methods as demonstrated by Fitzgibbon et al. (1996)\cite{fitzgibbon} but with severe drawbacks:
\begin{itemize}
\item A complex inversion process is required to compute the parameters.
\item Perfect fringes are required, since any flux, transmission or contrast variation during the fringe scan can bias the result by up to several degrees.
\item A moderately high success rate and stability is achieved in simulations and real data when compared to a classical $\chi^2$-minimization.
\end{itemize}

In this basic model we include the laser power fluctuations during the measurements $P(t)$, which can be extracted from a Fourier transform of the ABCD modulation of our measurements. The measured transmission variation of the phase shifter used as delay line as a function of applied voltage lies within the frequency of the laser power fluctuations such that it is filtered out by P(t). Therefore, our final model is the following expression 
\begin{equation}
I_i(t) = P(t)\left[a_i + b_i\sin(\beta(t) + \alpha_i)\right] \label{eq:final} \hspace{0.5cm} .
\end{equation}

With this function we determine the phase angles $\Delta_{AB}$ , $\Delta_{BC}$ and $\Delta_{CD}$ in our experiment. In more detail, we apply linear voltage ramps consisting of 1000 phase steps of about $\unit{4}{\nano\meter}$ to the phase shifter which is used as delay line. Due to the natural fringe drift this realization of a delay line is not limited by discrete voltage sampling. The ABCD modulation is run at frequencies close to $\unit{1}{\kilo\hertz}$. The fringe drift typically occurs at $\unit{0.1}{\nano\meter\per\milli\second}$ such that during an ABCD cycle the drift is smaller than $\unit{0.3}{\nano\meter}\simeq\lambda/6000$ and has no significant influence on our measurement of the ABCD phase angles.
\begin{figure}
   \begin{center}
   \begin{tabular}{c}
   \hspace{8.5mm} \includegraphics[height=7.8cm]{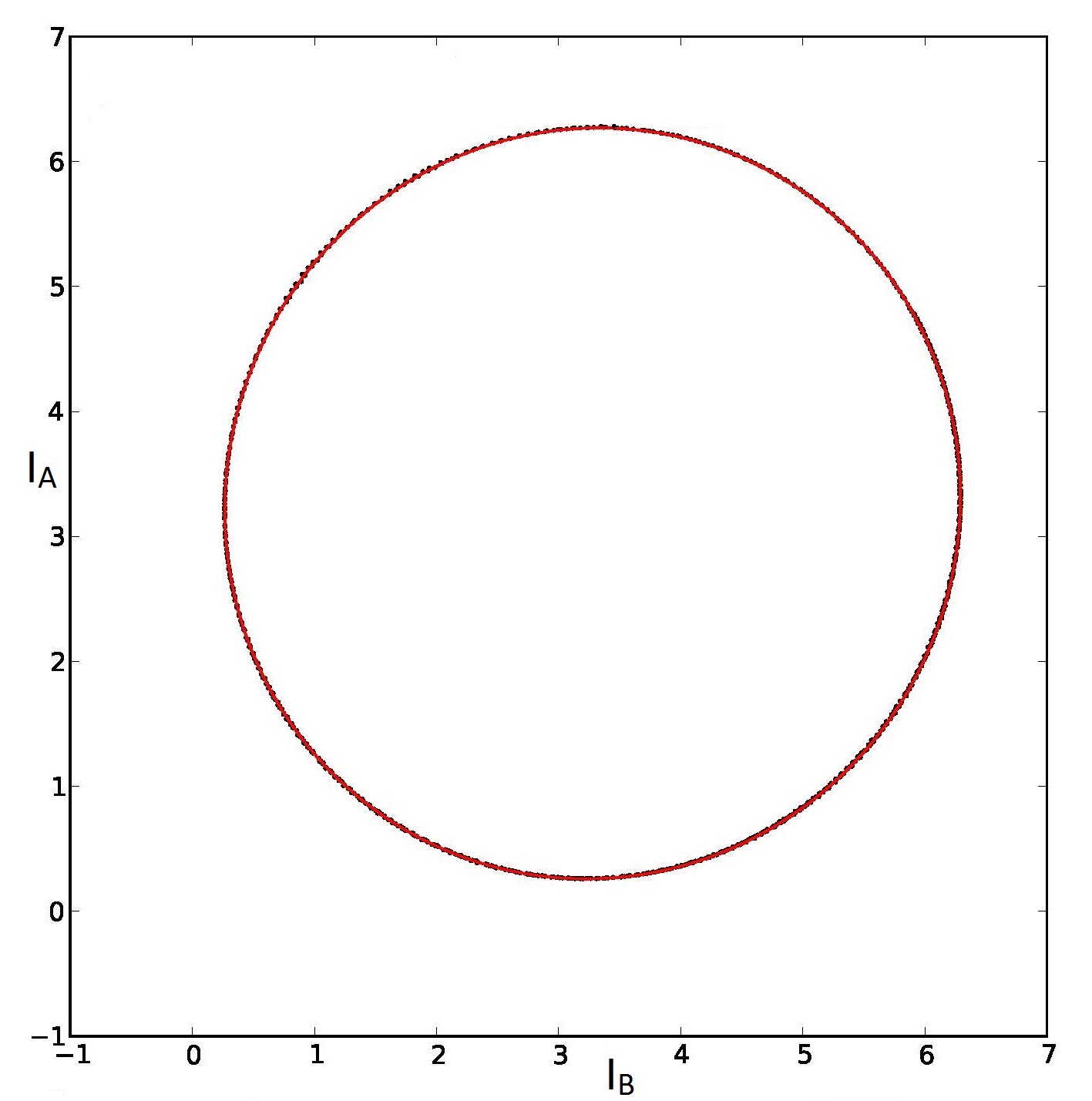} \\
   \hspace{0.0mm} \includegraphics[height=7.675cm]{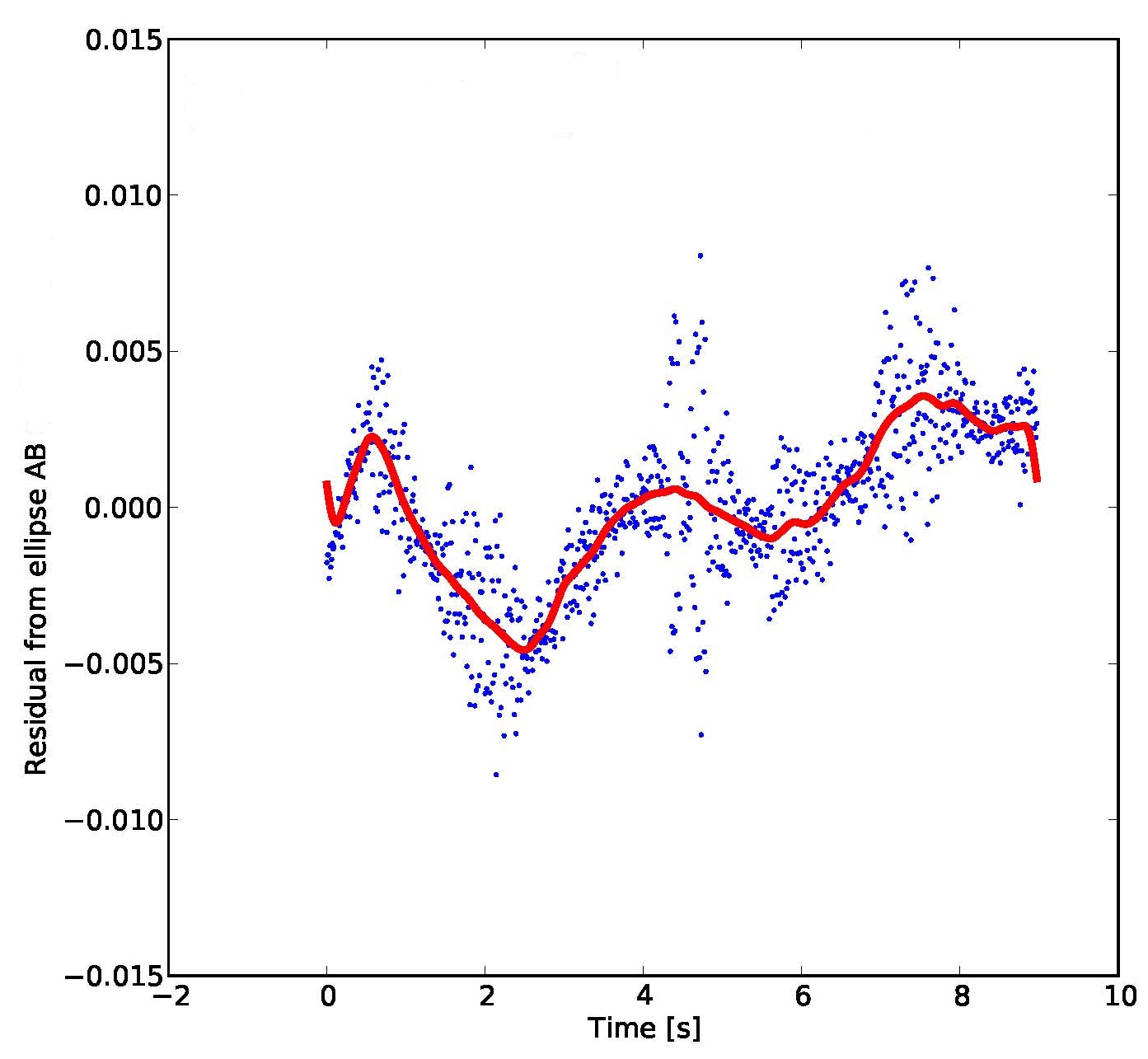} \\
   \hspace{1mm} \includegraphics[height=3.35cm]{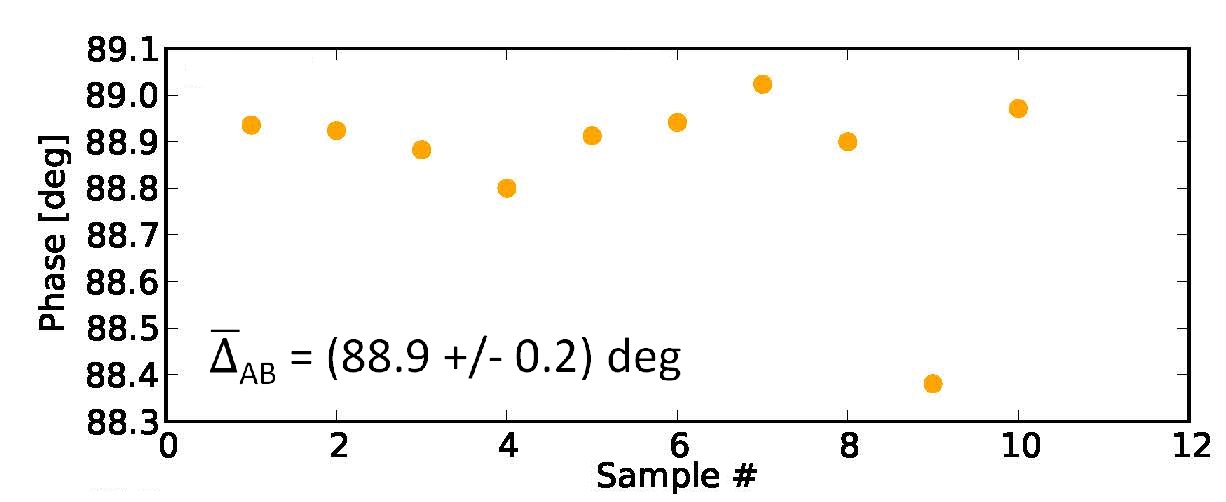}
   \end{tabular}
   \end{center}
   \caption[example] 
   { \label{fig:results} Performance of the phase-step insensitive algorithm. The top plot shows the fit of an ellipse to data acquired at a ABCD modulation of $\unit{700}{\hertz}$. The data are the voltages measured in Volt by the receiver and correspond to the intensities $I_A$ and $I_B$ for the phases A and B as a function of delay line shifts. The fit line is displayed in red and the black data points are almost invisible due to the small dispersion. Larger dispersions are found when the setup is perturbed by vibrations, but no bias is introduced since they average out. In the figure below the fitting residuals of the $\chi^2$-minimization relative to the fringe continuum are shown as a function of time. The bottom plot displays a series of 10 measurements with a mean phase step $\bar\Delta_{AB}=\unit{88.9}{\degree}\pm\unit{0.2}{\degree}$. }
   \end{figure} 

We took series of measurements at different ABCD frequencies in the range between \unit{0.5}{\kilo\hertz} and \unit{1.5}{\kilo\hertz} and different dates few months from each other without specific control of the environment to extract the phase shifter calibration error of this phase-step insensitive algorithm. Typically, the measured phase shifts are consistent within $\unit{0.2}{\degree}$ or $\unit{1}{\nano\meter}$  between the series which also is the typical dispersion on each individual series of measurements. These results validate that the phase-step insensitive algorithm is an appropriate method to calibrate the ABCD phase steps for the metrology on a $\unit{1}{\nano\meter}$-level. 

In comparison to the method of linearly scanning a full wave with the phase shifter, we not only achieve a precision of factor two higher, but we are also able to extract phase angles unambiguously. The advantage is that we do not need to model the interference fringes including the various complex effects that let the intensity distribution strongly differ compared to the theory of simple two-beam interference. The deficiencies of the linear scan method become obvious when directly comparing the results of both algorithms tested in Table \ref{tab:result}. In Figure \ref{fig:results} the determined phase steps $\Delta_{AB}$ of one such series are shown as well as one of the respective ellipses together with the corresponding fit and residuals.
\begin{table}[h]
\centering
\begin{tabular}{l|l|l|l}
\textbf{Results} & $\boldsymbol{\Delta_{AB}}$ & $\boldsymbol{\Delta_{BC}}$ & $\boldsymbol{\Delta_{CD}}$ \\ \hline
Applied voltage step & $\left\lbrace\unit{0}{\volt}, \unit{2}{\volt}\right\rbrace$ & $\left\lbrace\unit{2}{\volt}, \unit{4}{\volt}\right\rbrace$ & $\left\lbrace\unit{4}{\volt}, \unit{6}{\volt}\right\rbrace$\\ \hline 
Phase-step insensitive algorithm & $\unit{88.7}{\degree}\pm\unit{0.2}{\degree}$ & $\unit{89.2}{\degree}\pm\unit{0.2}{\degree}$ & $\unit{92.8}{\degree}\pm\unit{0.2}{\degree}$ \\ \hline 
Linear scan & $\unit{87.0}{\degree}\pm\unit{0.4}{\degree}$ & $\unit{86.6}{\degree}\pm\unit{0.4}{\degree}$ & $\unit{87.1}{\degree}\pm\unit{0.4}{\degree}$ \vspace{0.3cm}
\end{tabular}
 \caption[Results of the phase-step insensitive algorithm compared to the linear scan]{Results of the phase-step insensitive algorithm compared to the linear scan. For the phase-step insensitive method the average phase shift values $\Delta_{AB}$, $\Delta_{BC}$ and $\Delta_{CD}$ were obtained from series of measurements at different ABCD modulation frequencies and different dates during a time span of a few months. The absolute value of the applied voltage steps always is \unit{2}{\volt}, but results in different phase angles due to the non-linear behavior of the phase shifter. Obviously, the linear scan shows systematically different as well as less precise results and fails in revealing the non-linearities probably due to flaws of the respective fit model.}
\label{tab:result}
\end{table}

\section{CONCLUSIONS}

The calibration of the metrology phase shifter presented here fulfills the specified requirements for performing $\unit{10}{\micro\arcsec}$-level astrometry. We could demonstrate experimentally that ABCD phase shifts can be calibrated with an accuracy of typically $\unit{1}{\nano\meter}$ in a stable setup and environment. Despite non-linearities that occur in the component, the metrology fiber chain or in the environment we were able to elaborate a method that is not as sensitive to these as simple scans of fringes in order to extract the relation between the applied voltage and the resulting phase shift. The phase-step insensitive algorithm is able to determine phases unambiguously when a robust fitting routine for ellipses is included. In our case a $\chi^2$-minimization serves this purpose.
\newpage
These results were achieved under normal air pressure and room temperature. In GRAVITY the phase shifter is going to be operated in vacuum at $\unit{\power{10}{-6}}{\milli\bbar}$ cooled to $\unit{240}{\kelvin}$ -- conditions which might lead to a higher calibration accuracy provided that the instrument does not introduce higher noise levels. Currently, we analyze the instrument's behavior in this respect. Furthermore, fiber differential delay lines are implemented in GRAVITY which can be used for the calibration of phase shifts and do not show a transmission dependency as the phase shifter. However, due to the transmission modulation of the phase shifter we might need to adapt the original phase extraction formula of Equation \ref{eq:phi}, since the underlying fringe model of Equation \ref{eq:fringes} does not take this effect into account.

When the ABCD phase shifts will be calibrated to the specified accuracy within GRAVITY, an important step will be taken towards measuring the optical paths in the interferometer to a nm-level precision via phase-shifting interferometry and thus towards unprecedented astrometric errors at the level of $\unit{10}{\micro\arcsec}$.
\bibliography{report}   

\begin{thebibliography}{1}

\bibitem{glindemann}
Glindemann, A.,  [{\em Principles of Stellar
  Interferometry}{\nolinebreak\hspace{0.1em}]}, Springer-Verlag, Berlin
  Heidelberg (2011).

\bibitem{gillessen12}
Gillessen, S. et~al., ``{GRAVITY}: metrology,'' in [{\em Optical and Infrared
  Interferometry III}{\nolinebreak\hspace{0.1em}]},  Deplancke, F., ed., {\em
  Proc. SPIE} {\bf 8445},  84451O (2012).

\bibitem{sahlmann}
Sahlmann, J. et~al., ``The prima fringe sensor unit,'' {\em Astronomy {\&}
  Astrophysics}~{\bf 507}(3),  1739--1757 (2009).

\bibitem{wooten}
Wooten, E.~L. et~al., ``A review of lithium niobate modulators for fiber-optic
  communications systems,'' {\em IEEE Journal of Selected Topics in Quantum
  Electronics}~{\bf 6}(1),  69--82 (2000).

\bibitem{farrell}
Farrell, C.~T. and Player, M.~A., ``Phase-step insensitive algorithms for
  phase-shifting interferometry,'' {\em Measurement Science and
  Technology}~{\bf 5},  648--652 (1994).

\bibitem{fitzgibbon}
Fitzgibbon, A., Pilu, M., and Fisher, R.~B., ``Direct least square fitting of
  ellipses,'' {\em IEEE Transactions on Pattern Analysis and Machine
  Intelligence}~{\bf 21},  476--480 (1999).

\end{thebibliography}
\bibliographystyle{spiebib}   

\end{document}